\documentclass[12pt]{article}

\usepackage{epsfig,here}

\newcommand{\beq}{\begin{equation}}
\newcommand{\eeq}{\end{equation}}
\newcommand{\beqa}{\begin{eqnarray}}
\newcommand{\eeqa}{\end{eqnarray}}
\newcommand{\beqar}{\begin{eqnarray*}}
\newcommand{\eeqar}{\end{eqnarray*}}

\def \la {\langle}
\def \ra {\rangle}

\begin{document}

\title{\bf\Large
   Back-Reaction of Clocks and Limitations on Observability
       in  Closed Systems
}

\author{
Aharon Casher\footnote{\it ronyc@post.tau.ac.il} \
 and \ Benni Reznik\footnote{\it reznik@post.tau.ac.il}
{\ } \\
{ \small  School of Physics and Astronomy, Tel Aviv
University, Tel Aviv 69978, Israel}
         }

\date{September, 2, 1999}

\maketitle

\begin{abstract}
{ Measurements are ordinarily  described with respect to absolute
"Newtonian" time. In reality however, the switching-on of the
measuring device at the instance of the measurement requires a
timing device. Hence the classical time $t$ must be replaced by a
suitable quantum time variable $\tau$ of a physical clock. The
main issue raised in this article is that while doing so, we can
no longer neglect the {\em back-reaction} due to the measurement
on the clock. This back-reaction yields a bound on the accuracy of
the measurement. When this bound is violated the result of a
measurement is generally not an eigenvalue of the observable, and
furthermore, the state of the system after the measurement is
generally not a pure state. We argue that as a consequence, a
sub-class of observables in a closed system cannot be realized by
a measurement.}
\end{abstract}
\newpage
\section{Introduction}

The special role of time in quantum theory has been noted since
the early days of quantum mechanics. Unlike other observables time
remains a classical variable.  It cannot be a simply ``quantized''
because as is well known, for bounded Hamiltonians, a formal
construction of a time operator gives rise to a non self-adjoint
operator~\cite{pauli}. Nevertheless this difficulty becomes a
serious obstacle, only when a system must be quantized as a whole,
that is without first splitting between what is quantum and what
is classical. In this case,  since the total energy of a closed
system is conserved, all observables become independent of the
``external'' classical time. Time must therefore be constructed
``internally" from quantum variables alone. The conceptual as well
as technical difficulties in carrying out this task are sometime
referred to as `{\em The problem of time}' in quantum
theory~\cite{kuchar,hartle-clock,unruh-wald,page,englert,time}.

The aim of the present article is to examine the problem of time
from the point of view of von-Neumann's quantum measurement
theory~\cite{von-Neumann}. In this theory, von-Neumann handles both the system and
the measuring device on the same quantum mechanical footing, and
provides a framework consistent with the following fundamental
quantum postulates:

\begin{itemize}
\item[1.] {There is no fundamental restriction on the accuracy that
one can measure a {\em single} observable.}
\item[2.]{ The result
of a precise measurement is one of the eigenvalues of
the Hermitian operator representing the observable.}
\item[3.]{The state of the system after the measurement
is ``reduced" to the corresponding eigenstate\footnote{ In von
Neumann's framework the initial state of the system is not
(effectively or truly) reduced to the measured component as stated
in  2. and 3 above. It only shows that the correct correlation
between a measuring device observable and the system can in
principle be established. Issues concerning the interpretation or
the reality of the reduction are not considered in this article,
nor are they relevant to the conclusions pointed out here.}. A
precise measurement is hence also a preparation of a pure state.}
\end{itemize}

von Neumann's theory handles time as a
classical variable. In reality however the switching-on of the
measuring device at the instance of the measurement requires a
timing device. Hence the classical time $t$ must be replaced by a
suitable {\em quantum} time variable $\tau$ of a physical
clock~\cite{peres,hartle-clock}.
The main issue raised in this article is that
while doing so, we can no longer neglect the {\em back-reaction
due to the measurement on the clock}. We will argue that:

\begin{itemize}
\item[1'.]
{There is a fundamental restriction on the accuracy  we can
measure a single observable: 
\beq {\Delta J \over J } \ge {\hbar
\over (E_C-E_0) \delta T}     
\label{accuracybound}
\eeq 
}
\end {itemize}
Here, $\Delta J$ is the accuracy in
a measurement of $J$, $E_C$, the clock energy  is bounded from
bellow by $E_0$, and $\delta T$ is the duration of the
measurement.

We shall see that for a given clock energy, certain states of the
system will {\it only approximately}  satisfy postulates 2. and 3.
above. However for other states which evolve sufficiently fast
with respect to the time scale $\hbar/(E_C-E_0)$ the previous postulates are
clearly violated:

\begin{itemize}
\item [2'.]
{The result of a measurement is generally not
an eigenvalue of the observable.
}
\item [3'.]
{ The state
of the system after the measurement
is generally not an eigenstate of the observable,
nor will it be a pure state.}
\end{itemize}

In fact the purity of the state of the system after the measurement
is at best a good approximation.
Generally after reduction to a certain value
of the measuring device ``pointer'',
the system remains correlated with the clock, i.e. it
is in a mixed state.
If we further measure the state of the clock and by that
reduce the system into a pure state,
the state of the system will not be correlated with
the result of the measurement.

Why does von-Neumann's theory fail when the timing device is
quantized? There are two separate reasons for this  breakdown.
First, due to the time-energy uncertainty relation, there is a
minimal inaccuracy, and the time of the measurement becomes
"fuzzy" by $\Delta \tau  \ge \hbar/\Delta E_C $. Consequently,
even if we can measure with arbitrary accuracy,  the time of the
measurement is still unknown by $\Delta \tau$. If during this time
the system evolves considerably the final state of the system
after the measurement is still correlated with the clock.

The second feature is the back-reaction on the clock
\cite{peres,arrival,weighing}. When the measuring device becomes
too accurate, i.e. $\Delta J/J< 1/E_C\delta T$, the back-reaction on
the clock increases and the clock's ``trajectory'' is distorted.
It no longer shows the correct interaction time. For even better
accuracy the clock is mostly ``reflected back"  in its coordinate
time $\tau$.  The phenomena is reminiscent to unitarity violations
found in the WKB approach to time in quantum
gravity\cite{banks,casher-englert}. We will show that even for the
minor part of measurements in which the clock is not reflected,
the result still may not be interpreted as a successful measurement,
since the required correlations between the measuring device and
the relevant observable's eigenvalues are absent.

We will proceed as follows:
In section 2. we first consider a measurement with an ideal
 timing device, i.e. the case that the clock Hamiltonian is unbounded.
 In this case the back-reaction does not effect the measurement.
We show that observables defined "relative" to the clock time
commute with the total Hamiltonian, in accordance to the
``evolving constants of motion'' approach of
Rovelli~\cite{rovelli1,rovelli2}. In section 3. we replace the
ideal clock by a physical clock  and derive  the consequent bound
(\ref{accuracybound}). 
An
exactly solvable toy model is used in section 4. to examine in
some details the  breakdown of von Neumann's theory in the
impulsive limit. In section 5. we conclude with some general
remarks. In the appendix we review the basic properties of the ``free 
particle clock''  used in sections 3 and 4. Throughout the article we choose units such
that $\hbar=1$.

\section{Ideal Clocks in a Closed System}

The main issue of the article is to examine the effects of timing
measurements by physical clocks.
To relate our approach and result to the work already done on the
"problem of time", we begin by first considering the
case of a quantum measurement with an ideal clock in a closed system.
By an ideal clock we refer to the unphysical case of a clock with
unbounded energy.
If for instance the Hamiltonian of the clock is linear
\beq
H_C = cP_x
\eeq
the coordinate $x$ can be used to define the clock time $\tau=x/c$.
As we now show, the measurement problem,  can be in this case
well formulated relative to the clock time $\tau$.

Consider a closed system which includes an ideal clock and a
a measuring device, and consider a measurement of some observable
$J$ at some internal time $\tau=\tau_0$.
The full Hamiltonian is
\beq
H_{Closed}=H_C + H_{MD}(Q,P)+ H(J,\beta) + g(\tau-\tau_0)QJ
\label{Htotal}
\eeq
where $\beta$ stands for other degrees of freedom of the system.
For an ideal clock, $\tau$ is conjugate to the clock Hamiltonian
which may be represented as $H_C = -i \partial/\partial\tau$.
The last term describes a standard von-Neumann
measurement of an observable $J$
at the clock time $\tau=\tau_0$.
For simplicity, and without loosing generality, we will next 
take $H_{MD}$, the Hamiltonian of the measuring device,
be arbitrarily small.

Since $H_{Closed}$ is independent of the external time $t$, and
since the system is closed, we may as well assume the closed
system  to be in an eigenstate of energy $E=E_0$, 
\beq H_{Closed}
|\Psi\ra = E_0|\Psi\ra \eeq As  function of $\tau$ we get \beq
|\Psi(\tau) \ra = \exp\biggl( -i(\tau-\tau_i)E_0 -i
Q\int_{\tau_i}^\tau g(\tau') J d\tau' \biggr) |\Psi(\tau_i)\ra 
\eeq 
Let us
compare the correlations between the measuring device pointer variable, $P$,
and the observable $J$ when  $\tau_i<\tau_0$ and when $
\tau>\tau_0$. In the impulsive limit $g\to \delta(\tau-\tau_0)$
the dynamical effect of $H(J,\beta)$ can be neglected and the
measuring device pointer  is shifted when $\tau>\tau_0$
by the value of $J$. Thus, if for $\tau<\tau_0$, there are no
correlation between the measuring device and the system, for
$\tau>\tau_0$ the correlations  are  of the same
form as in an ordinary von-Neumann measurement. We also note that
the shift in $P$ is independent of the particular total energy
$E_0$ which affects only the overall phase.

It is often stated that non-trivial time dependent observables
cannot be measured in a closed system. The basic argument goes
like that: The Hilbert space of states describing a closed system
is made of all states which are degenerate eigenstates of the
total Hamiltonian. By the degeneracy  any general
state in this Hilbert space is also time independent. 
Therefore, only time independent operators, which commute with
$H_{Closed}$, constitute the set of  observables for a closed system.
Notice that this argument is valid for a bounded as well as
unbounded Hamiltonian.

In our example above, since $H(J,\beta)$ is arbitrary,
and since $\tau$ is conjugate to $H_C$ we have
\beq
[J, H_{Closed}] = [J, H(J,\beta)] \neq 0
\eeq
and
\beq
[\tau, H_{Closed}] =i
\eeq
Hence both the clock-time $\tau$ and the observable that we measure
are according to this no-go theorem, non measurable observables!

To resolve this apparent contradiction we notice that in a closed
system, only {\it relative} observables may be measured \cite{as}.
As is well-known, observables
as position, velocity, angular momentum, etc, both in classical
mechanics as well as in quantum mechanics, are {\it relative}
observables. Indeed, we never measure the absolute position of a
particle, but the distance in between the particle and some other
object. The relative location,
 $r=x_1-x_2$ of two particles does
commute with the total momentum $p=p_1+p_2+\dots$.
Similarly, we never measure the angular momentum of a
particle along an absolute axis, but along a direction defined by
some other physical objects. Therefore the angular momentum of a
closed system can be measured only with respect to a point within
the system, say the location of the center of mass, and along a
direction defined by constitutes of the system.
The absolute value of an observable is meaningless, and we have
to relate this value to a `reference system'.

In analogy, a time dependent observable, $J(t)$, cannot be
measured at time $t$, because $J(t)$ does not commute with the
Hamiltonian. However, we argue that any  observable may be
measured at an arbitrary time defined by means of an
internal clock.

Suppose as an example we take $J=X$ of a free particle with $H=P_X^2/2$,
and consider a measurement at internal time  $\tau_0$.
The observable we measure is thus
\beq
\hat X(\tau_0) = \hat X - \hat P(\hat \tau -\tau_0)
\label {constant}
\eeq
To avoid confusion, for the rest of this section we will add to
operators a `hat', therefore $\hat \tau$ is an operator, while
 $\tau_0$ is a real number.
Since the value of $\hat X$ at a certain instance does not change
in time indeed
\beq [\hat X(\tau_0), H_{Closed}]=0
\eeq
Likewise,
from the point of view of an external observer, $\hat X(\tau_0)$
is again a constant of motion that can be measured at any external
time $t$. The internal clock time, $\hat \tau$, and $t$ are
linearly related by $\hat\tau(t) = t + \hat \tau(0)$. 
Therefore, the external time $\hat t_0$ of the measurement is $\hat t_0
= \tau_0 - \hat\tau(0) = t -(\hat \tau(t) - \tau_0)$
Using the latter relation we can re-express $X(\tau_0)$
in  terms of the  external time $t$ as 
\beq \hat X(\tau_0)= \hat X(\hat t_0)= \hat
x(t) - \hat p(\hat \tau-\tau_0)
\label{tconstant}
\eeq
Since $[x(t) - \hat p(\hat \tau-\tau_0),H_{Closed}]=0$
this is again a constant of motion. Notice that
since  $\hat t_0$
is an  operator the  time of the measurement  remains uncertain 
with respect to $t$.

To summarized, our model suggests that the ``time dependent''
observables are in fact constant of motions. They are measured
with respect to physical clock time and correspond to eigenstates
of ``composed'' clock-system operators as in eq. (\ref{tconstant}).
More generally the observables are of the form 
\beq \hat J(\tau_0)
= U( \hat  \tau- \tau_0) \hat J(t)U^\dagger(\hat
\tau-\tau_0) \eeq For each real number  $\tau_0$ we obtain another
constant of motion.

Our discussion  above seems to support the view  suggested by
Rovelli \cite{rovelli1,rovelli2}. According to him, the evolution
in time of an observable is pictured as ``evolving constants of
motion". Indeed, in our case,  a sequel of measurements of the
same operator $A$, at different clock times $\tau_{0i}$, yields a
correlation between the ``pointer'' coordinate $\hat P_i$, of
measuring device $i$, and the corresponding constants of motions
$\hat J(\tau_i)$.

\section{Back-Reaction}

By assuming that clock Hamiltonian is unbounded, we have avoided
the possibility of back-reaction on the clock due to the
measurement. In the previous  section, the trajectory of the time
variable $\tau$ was linearly correlated to the external time
irrespective of the coupling function $g(\tau)$. Since stable
physical systems must have a ground state, we  next reformulate
the measurement problem using clocks with bounded energy. The
general properties of physical clocks are described in the
Appendix, where we also present the ``free particle'' toy model
for a clock with a Hamiltonian $H_C=P_X^2/2M$. If the average
momentum of the particle is much larger than the momentum
uncertainty, $\la P_X\ra \gg \Delta P_X$, we can use as an
approximate time variable the operator 
\beq \tau \equiv {X\over
{\la P_X\ra /M}}, \ \ \ \ \ \ [\tau, H_C]= i +O\biggl({\Delta P_X\over
\la P_X\ra }\biggr)
 \label{taudef} 
\eeq

A Hamiltonian describing a measurement of an observable $J$,
is hence
\beq
H = {P_X^2\over 2M} + H(J,\beta) + g(X)QJ
\eeq

The coupling function, $g(X)$, will be taken to be non-vanishing
only in a finite region $-X_0/2< X < X_0/2$.

\begin{figure}[H]
  \begin{center}
\mbox{\kern-0.5cm
\epsfig{file=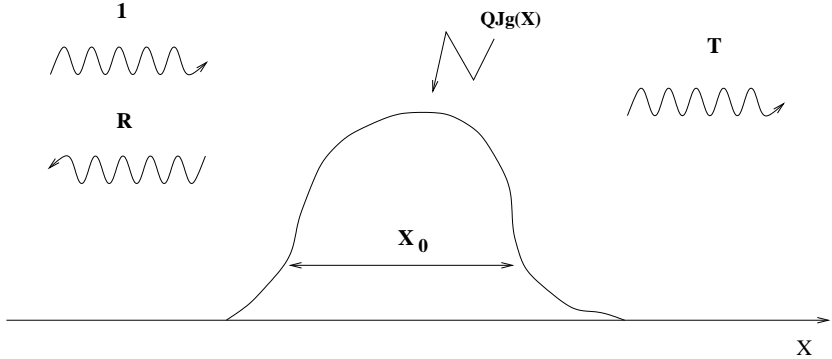,width=11.0true cm,angle=0}}
\label{figure1}
  \end{center}
\caption{ The effective potential of the clock, $Qjg(X)$, is
plotted as a function of the ``time'' coordinate $X$. At $X<X_0$
we must include back scattered (left moving) clock states.  }
\end{figure}

There are two sources which contribute to the
the uncertainty in the time of the measurement.
The first is the quantum uncertainty $\Delta \tau$
(eq. (\ref{dtau}) in the appendix).
This quantum fuzziness of time is determined by the
time-energy uncertainty relation
\beq
\Delta \tau \Delta E \ge 1
\eeq
where $\Delta E$ is the uncertainty in the clock kinetic energy, and
$\Delta \tau$ is the uncertainty associated with  the uncertainty in the
clock coordinate $X$.

The other source of uncertainty is due to the non-impulsive nature
of $g(X)$. We will denote the latter by $\delta T$.
  If the relation (\ref{taudef}) between $\tau$ an $X$ is still
respected, we have
\beq
\delta T \simeq {X_0 \over (\la P_X\ra /M)}
\eeq
However "during" the measurement (i.e. for $X\in (-X_0/2,X_0/2)$),
the relation between $X$
and $t$ may be modified and the latter
equation should be modified accordingly.

Quantum and classical measurements differ in one important
respect. Only classically can we make the disturbance to the
system arbitrarily small. In particular, since  the accuracy of the
measurement  is proportional to $\Delta P \approx 1/Q$, the
strength of the interaction always scales like $\sim 1/\Delta J$,
i.e. it diverges for $\Delta J\to 0$.

In our case this interaction also affects the clock. According to
the interpretation of $X$ as proportional to time, initially the
clock wave function must be localized at $X\ll 0$ and after the
measurement at $X\gg0$. Therefore a "successful" measurement can
be viewed as a scattering experiment for the clock, with only little
back-scattered waves. However, an obvious consequence of the
discussion above, it that in the limiting case $\Delta J\to 0$,
the clock wave function will be mostly reflected back.  Unlike the
ordinary case where the accuracy determines only the disturbance
caused to non-commuting observables, in our case the measurement
affects the chance that an event of a measurement actually
happens. A related "back-scattering" effect was previously
discussed for in the case of a measurement of time of arrival
\cite{allcock,time,arrival}.

To obtain a qualitative relation between the relevant scales in
the problem: the accuracy of the measurement, the clock energy and
accuracy, and the duration of the measurement let us examine the
condition for very small back reaction. A sufficient requirement
is that the kinetic energy $E_C$ of the clock be large enough with
respect to the interaction term
 \beq 
E_{C} \gg {g_0} Q J
\label{largeE} 
\eeq
where $g_0$ is the average value of $g(X)$.
This ensures that back-scattering effect is small and that during
the measurement the clock's motion remains free up to small
corrections. Therefore, up to corrections of order $O(\Delta
P_X/\la P_X\ra )$,
 the eigenvalue equation $H\Psi =0$ for the closed system
becomes
\beq i{\partial \Psi\over \partial \tau} = \biggl(
H(J,\beta)+ g(X(\tau)QJ \biggr) \Psi
\eeq
where $\tau$ is given by
(\ref{taudef}).
Hence the shift of the pointer variable is \beq
P_{MD}(\tau\gg0) - P_{MD}(\tau\ll 0)
 = \int J g(X(\tau) )d\tau \approx g_0X_0{M\over \la  P_X \ra} \bar J
\eeq
were $\bar J= \int gJd\tau/\int gd\tau$, is the time average of $J$ over the
time $\delta T$.
We can now obtain a relation between the
 $\Delta P_{MD}$ of the ``read out'' measuring device coordinate
and the accuracy $\Delta J$ of the measurement 
\beq 
\Delta P_{MD}
\simeq g_0X_0 {M\over \la P_X \ra }  \Delta J \biggl( 1 + {\bar
J\over \Delta \bar J}{\Delta P_X\over \la P_X\ra
  }\biggr)
  \label{pmd}
\eeq 
The last term, due to the uncertainty in $P_X$, contributes
an additional uncertainty to the measurement. It may be
interpreted as resulting from the uncertainty in the time of the
measurement $\Delta \tau$. If the back-reaction is small, as in our
case, this error is controlled by the time-energy uncertainty
relation.  As will be shown bellow we may neglect this term by
taking sufficiently large $\la P_X\ra $.

 Therefore we obtain the relation
 \beq
 \Delta Q \ge {\la
P_X\ra \over g_0 X_0 M}{1\over \Delta \bar J} \label{deltaq}
 \eeq
 Since we wish to minimize the right hand side of
(\ref{largeE}), we set $\la Q\ra =0$.  We hence obtain the
inequality
\beq
{\Delta \bar J \over \bar J} \gg {1\over E_{C}
\delta T}\approx {1\over X_0\la P_X\ra } \label{bound}
\eeq

For measurements with accuracy $\Delta J/J = \epsilon$, the
duration of the measurement must satisfy $\delta T\gg (\epsilon
E_C)^{-1}$. The exact limit $\epsilon\to 0$ can be satisfied only
with $E_C\to \infty$. We further note that the above inequality
implies that ${J\over\Delta J}{\Delta P_X\over \la P_X\ra} \ll X_0
\la P_X\ra$. Hence the approximation used in obtaining
(\ref{deltaq}) is justified in for sufficiently large $\la
P_X\ra$.

 The inequality (\ref{bound}) is satisfied for time averaged
observables. Therefore, the measured value will determine the
state of the system as one of the eigenvalues only if $J$ is
nearly constant during the time $\delta T$. Otherwise  after the
measurement the system remains in a mixed state. Let us denote by
$\omega$ the typical frequency of $J$ ($\omega \approx
\la[J,H(J,\beta)]\ra/\la J\ra$) . We hence need to satisfy 
\beq
  1/\omega \gg \delta T \gg 1/E_{C}
\eeq
or $\omega \ll E_{C}$. Therefore,  only if $\omega$ is
much smaller then the clock frequency $1/E_{C}$,  can we
determine the state of the system after the measurement. It
follows that the state of the subsystem with the highest energy
available can not be measured.

\section{The impulsive limit}

In non-relativistic
quantum mechanics there is no fundamental restriction on the
duration, $\delta T$, of a measurement. The possibility of
taking the limit $\delta T\to 0$ implies in principle
the measurability at an instant of time.
In the previous section we have argued that a necessary condition
for an accurate measurement
is that $E_C\delta T\gg1 $.
We shall now examine in more details the consequences
when this inequality is violated.

For concreteness we consider a solvable model for a measurement
described, as before,  by the interaction
\beq g(X)JQ \eeq
where
the observable $J$ is 
for simplicity we taken commute with the
Hamiltonian $H(J,\beta)$. The coupling function is \beq g(X) =
{\lambda\over X_0}(\theta(X-{X_0\over2}) -\theta(X+{X_0\over2}))
\eeq and the clock Hamiltonian is taken as before to be a free
particle of mass $M$.

\begin{figure}[H]
  \begin{center}
\mbox{\kern-0.5cm
\epsfig{file=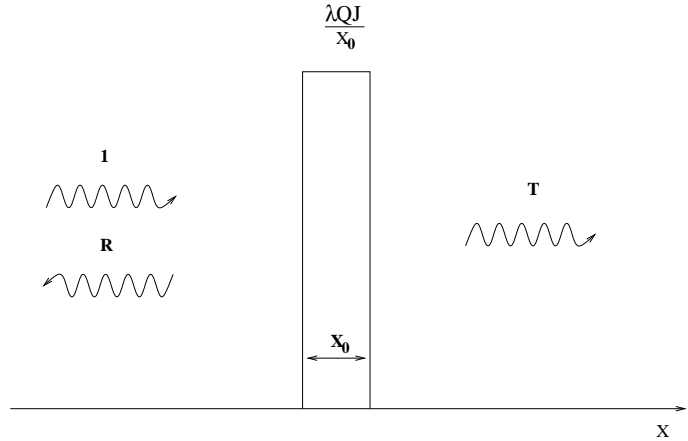,width=9.0true cm,angle=0}}
\label{fig2}
  \end{center}
\caption{The effective potential in the impulsive limit $X_0\to
0$, $\int g(X)dX =\lambda$  }
\end{figure}

Let the initial state of the system be a direct product of 
the clock state $|C\ra$, the measuring device state $|MD\ra$
and the state of the system $|S\ra$:
\beq 
|\Psi\ra_I =
|C\ra |MD\ra|S\ra
\eeq 
for $X<0$. 
Then the
transmitted wave function at $X>0$ is given by 
\beq 
|\Psi\ra_T =
\sum_jC_j\int dk \phi_C(k) T(k,Q,j)) e^{ikX} |k\ra |MD\ra|j\ra 
\label{psit} 
\eeq 
Here $\phi_C(k)=\la k|C\ra$ is the clocks wave-function in the 
free momentum representation $k^2=2ME_C$, 
and $C_j= \la j|S\ra$.

In the present case, the clock rate is affected during the
interaction. 
For $|X|<X_0$, the momentum is given by 
$q^2=2M(E^0_C + \lambda Qj/X_0)$. For sufficiently
large $Q/X_0$ this momentum is completely dominated by the
interaction rather than by the free momentum $k$. 
Hence the duration
of the interaction is now 
\beq 
\delta T \simeq {MX_0/q} 
\eeq The
condition (\ref{bound}) will therefore be violated if 
\beq
E_C\delta T = q X_0 < 1 
\eeq 
where $E_C=q^2/2M$. The limit
$qX_0\ll1 $,  corresponds to the impulsive limit. In this case the
analysis is greatly simplified. The transmission amplitude becomes
\beq T(k) 
\simeq {1\over 1+ i \alpha Q}
 \ \ \ \alpha \equiv {\lambda j M\over k}
\eeq

In this limit we can therefore evaluate the shift of the measuring
device pointer variable $P$ for a the case of transmitted clock
state. Let us set for the initial state  of the measuring device 
in (\ref{psit}), 
a Gaussian wave $\la Q|MD\ra\equiv \chi_{MD}(Q) = exp(-Q^2\Delta^2/4)$.
The accuracy of the measurement is therefore given by $\Delta$.
The final wave function of the measuring device in the pointer representation
can be evaluated to  
\beqa 
\chi_{MD}(P)= \int dQ e^{iQP} {1\over
1+i \alpha Q} e^{-Q^2\Delta^2/4} = \nonumber \\
 {1\over \sqrt4\alpha}\biggl (1-
{\rm erf}\biggl({1\over 2}{\Delta^2-2\alpha P\over
\Delta \alpha} \biggr) \biggr )
\exp
\biggl(- {P\over\alpha} +  {\Delta^2\over 4\alpha^2}\biggr)
\eeqa

To clarify the meaning of the last equation let us examine the
final state of the measuring device in two limiting cases: when
the disturbance caused to the clock wave function is small, i.e.
the transmission amplitudes is nearly unity $T\simeq 1$, and the
second case that of a strong back-reaction on the clock: $T\ll 1$

In the first case $|T|\simeq 1 $ requires 
\beq {g_0 j QM\over k} =
\alpha Q \approx {\alpha\over \Delta} \ll1 \label{weak} 
\eeq
Therefore we have 
\beq T\simeq  \exp(i\alpha Q) \eeq 
Substituting
$T$ back into (\ref{psit}) we find that the
 shift caused to the measuring device pointer is
\beq \delta P_{MD} = \alpha = \lambda {M\over k} j \eeq 
It is
indeed proportional to the observable eigenvalues $j$. However, since by the
uncertainty relation
 $Q\sim \Delta Q \ge 1/\Delta P_{MD}$,
the "weakness" condition
(\ref{weak}) restricts the accuracy and yields
\beq
{\Delta j \over j} \gg 1
\eeq

\begin{figure}[H]
  \begin{center}
\mbox{\kern-0.5cm
\epsfig{file=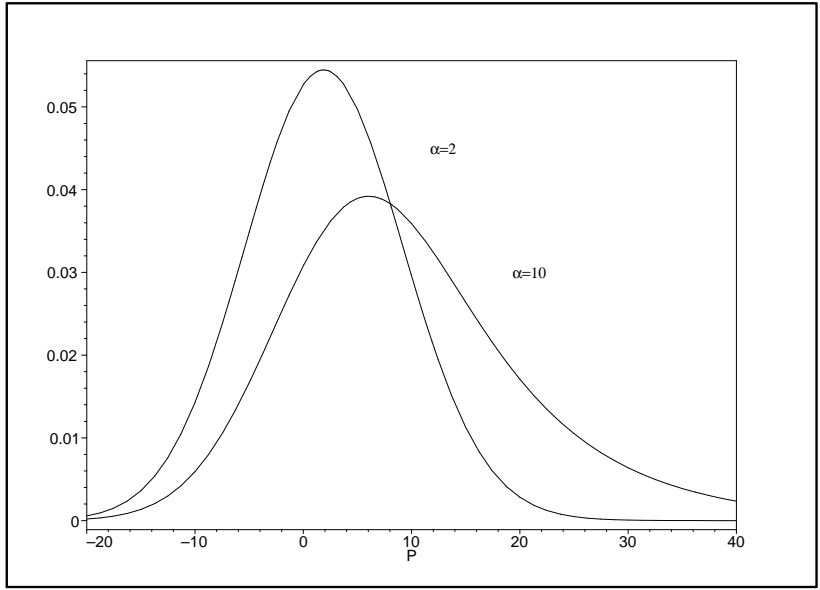, width=8.0true cm,angle=-90}}
\label{figure3}
  \end{center}
\caption{The pointer final wave function for the case of weak
back-reaction. The initial pointer wave function is a Gaussian of
width $\Delta=10$ centered around $P=0$. The shifts in the two
cases here are $\alpha =2$ and $\alpha=10$. In the latter case the
non-symmetric distortion indicates an  increase back-reaction.  }
\end{figure}

The behavior of the pointer in the first weak limit
is exhibited in Fig. 3.
 for different values of $\alpha<\Delta$.
In this weak limit  a single measurement is unable to resolve
between the observable eigenvalues. When the measurement is
repeated many times the average shift will tend to the expectation
value of $J$. However in each trail we cannot distinguish between
different eigenvalues.

\begin{figure}[H]
  \begin{center}
\mbox{\kern-0.5cm
\epsfig{file=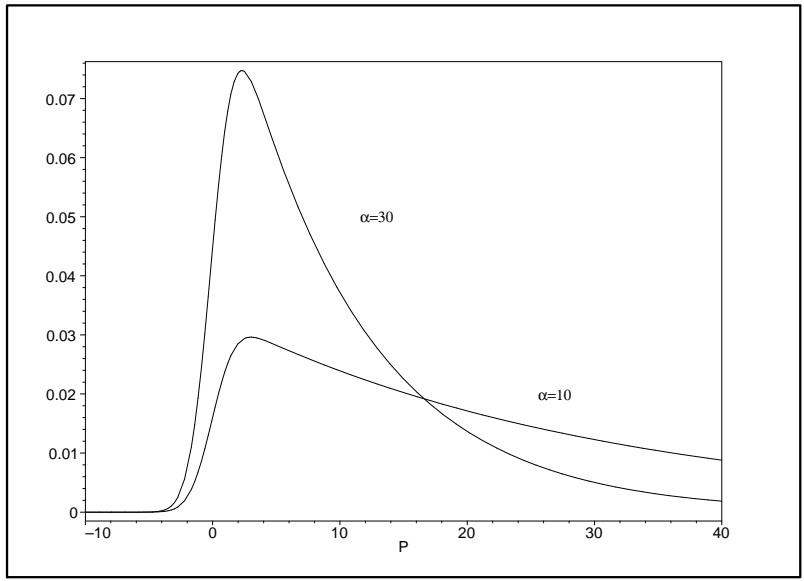,width=8.0true cm,angle=-90}}
\label{figure4}
  \end{center}
\caption{The pointer`s final wave function for the case of strong
back-reaction, occurring when the measurement becomes accurate
$\Delta < \alpha$. Here the resolution is fixed as $\Delta=5$ and
$\alpha = 10,20,30$}
\end{figure}

Next let us examine the ``strong'' limit, exhibited in Fig. 4.,
when due to the back-reaction the major  part of the clock wave
function is reflected back and $|T|\ll 1$. This corresponds to 
the case case
$\alpha\gg \Delta $, of high resolution. We find 
\beq
 |\Psi\ra_T \simeq
\int dk\phi_C(k)  \sum_{j} C_j e^{-{k\over M}\big|{P\over j}\bigr| }
\theta(Pj)|k\ra|P\ra|j\ra
\eeq

If for instance we find $P=P_0>0$ this reduces (``after'' the
measurement) to 
\beq 
|P_0\ra\sum_{j>0} C_j |\phi_j\ra|j\ra 
\eeq
where 
\beq 
|\phi_j\ra = \int dk e^{-{k\over M}\bigl|{P_0\over
j}\bigr|} \phi_C(k)|k\ra 
\eeq 
The state of the system  hence remains in a mixed state.
Notice that the states $|\phi_j\ra$ are  $not$ an orthogonal set of 
clock states.
Hence by further measuring (post-selecting) the state  of the
clock we cannot reduce the system into one of the eigenstates $j$.

To summarize, in the impulsive limit, none of the cases examined above
can realize an ordinary  measurement.  
In the weak limit, when the back reaction is small, 
we can observe statistically the expectation value of the observable, 
but cannot resolve eigenvalues. 
In the strong back-reaction limit, corresponding to an accurate
measurement, again we did not resolve the eigenvalues, and the final 
state of the system remains entangled with that of the clock.
A  further measurement on the clock
does not help to reduce the system to a definite pure state.

\section{Discussion}

The harmony between the fundamental postulates of quantum
mechanics and von-Neumann's measurement theory, breaks down if a
physical clock, with bounded energy, is used as a timing device
for the measurement. We have seen that to avoid back-reaction the
accuracy of the measurement must be limited and is bounded by
$\hbar/ E_C\Delta T$. Hence, the energy of the clock sets up a
scale of minimal time $\hbar /E_C$. Only for observables which
evolve slowly  in that time does the ordinary the theory  of
measurement apply with small corrections. For other cases  
a measurement does not ``reduce'' the state to a pure state
corresponding to an eigenvalue of the observable. Therefore the
state of a sub-system of a closed system, may not be observed in
the usual sense. Perhaps this indicates that the quantization of
closed systems must incorporate mixed states on a fundamental
level.

\vspace{2cm}

 {\bf Acknowledgment}
We thank Yakir Aharonov for many helpful discussions. B. R.
acknowledges the support from grant 614/95 of the Israel Science
Foundation, established by the Israel Academy of Sciences and
Humanities. \eject

\appendix
\section{Appendix: Free particle clock}
In this appendix we will review the basic properties of
the free particle toy model used as a clock in this article, 
with the Hamiltonian
\beq
H_C = {P_X^2\over 2M}
\eeq
as is well known one cannot construct an exact (conjugate)
time operator for a bounded
Hamiltonian which is also self adjoint.
Instead we consider an  approximate time operator 
define by
\beq
\tau = {X\over (\la P_X\ra /M)}
\label{tau}
\eeq
where $\la P_X\ra$ is the expectation value of $P_X$.
We have
\beq
[\tau, H_C] = i {P\over \la P_X\ra}= i + O\biggl(i{\Delta P_X\over
  \la P_X \ra} \biggr)
\label{tau,H}
\eeq
If the clock state are chosen to satisfy $\Delta P/P \ll1$
the right hand side can be made close to $i$.

Let us examine  the requirements needed
from a clock for a specific initial Gaussian state.
The degree to which this clock can approximate an ideal-clock
is defined by its accuracy and by the duration it is usable.
The accuracy of the clock is defined by how close can we
approximate the ideal Newtonian time $t$, by the physical time $\tau$.
The usable time, is determined by the maximal time interval
that the uncertainty $\Delta \tau$ does not grow beyond the
desired accuracy.

For the free particle clock,
$\tau$ deviates from the ideal time $t$ by the uncertainty
\beq
\Delta \tau \approx {\Delta X(t)
\over (\la P_X\ra/M)}
\approx  {\Delta X(0) \over (\la P_X\ra/M)}\sqrt{ 1 + \tau^2/M^2\Delta X^4(0)}
\label{dtau}
\eeq
The usable time interval is hence
\beq
\tau_{usable} \approx M (\Delta X)^2
\eeq
or
\beq
{\tau_{usable}\over \Delta\tau} \approx {\bar P^2\over M} \Delta \tau
\approx {\bar E \over \Delta E}
\eeq
Obviously we would like the left hand side to be much larger than
unity.
As we see this can be obtained for states with
$E\gg \Delta E$, or alternatively, this implies  $\bar P \gg \Delta P$
and $\Delta X \gg 1/P = \lambda_{de-Brolie}$.
The latter conditions are identical to the conditions following
from equation (\ref{tau,H})

\end{document}